\documentclass[11pt,a4paper]{article}

\usepackage{amsmath}
\usepackage{amssymb}
\usepackage{graphicx}
\graphicspath{{./}}
\usepackage{url}

\makeatletter
\newcommand{\affiliation}[1]{\gdef\@affiliation{#1}}
\newcommand{\emailAdd}[1]{\gdef\@email{#1}}
\renewcommand{\abstract}[1]{\gdef\@abstract{#1}}
\newcommand{\keywords}[1]{\gdef\@keywords{#1}}
\gdef\@affiliation{}
\gdef\@email{}
\gdef\@abstract{}
\gdef\@keywords{}
\renewcommand{\maketitle}{%
  \begin{center}
    {\Large\bfseries \@title\par}
    \vspace{0.8em}
    {\@author\par}
    \vspace{0.5em}
    {\@affiliation\par}
    \vspace{0.5em}
    {\ttfamily \@email\par}
  \end{center}
  \vspace{1em}
  \noindent\textbf{Abstract.} \@abstract\par
  \vspace{0.8em}
  \noindent\textbf{Keywords:} \@keywords\par
  \vspace{1.2em}
}
\makeatother

\newcommand{\safeincludegraphics}[2][]{%
  \IfFileExists{#2}{%
    \includegraphics[#1]{#2}%
  }{%
    \fbox{%
      \begin{minipage}[c][0.22\textheight][c]{0.78\textwidth}
        \centering
        Missing figure file:\\[0.5em]
        \texttt{\detokenize{#2}}
      \end{minipage}%
    }%
  }%
}

\title{\boldmath Scalar–Photon Coupling and Modified Photon Evolution: CMB Signatures and Cosmological Constraints}

\author{Yousef Bisabr}
\affiliation{Department of Physics, Shahid Rajaee Teacher Training University,\\
Lavizan, Tehran 16788, Iran}
\emailAdd{y-bisabr@sru.ac.ir}

\abstract{We investigate the constant-$\epsilon$ limit of a photon-sector modification
derived from an early-time interaction between a scalar field and radiation.
Starting from the underlying scalar--photon coupling, the parameter
$\epsilon$ quantifies the departure of the photon energy density from the
standard adiabatic scaling, leading in the constant-$\epsilon$ limit to
$\rho_{\gamma}\propto a^{-4+\epsilon}$ and to the modified CMB
temperature--redshift relation
$T(z)=T_{0}(1+z)^{1-\epsilon/4}$.
The standard photon sector is recovered for $\epsilon=0$.

We implement the corresponding constant-$\epsilon$ background evolution in
CLASS and verify that the numerical photon-density evolution reproduces the
analytic scaling. We then examine the resulting changes in the recombination
history, visibility function and CMB temperature-anisotropy spectrum.
These CMB spectra are used only as diagnostic outputs since the coupled
scalar--photon perturbation equations are not implemented self-consistently
in the numerical calculation and a full CMB likelihood analysis is beyond
the scope of the present work.

As a preliminary statistical application, we interface the modified CLASS
implementation with MontePython and analyze the Pantheon+SH0ES supernova
likelihood together with BAO measurements. Within the adopted
constant-$\epsilon$ framework, prior choices, data combination and fixed
early Universe parameter setup, we obtain
$\epsilon=0.0230\pm0.0065$.
The posterior is therefore centered on a positive value within this restricted
analysis. Because the BAO predictions depend on the model-dependent drag
scale and several early-Universe quantities are held fixed, this result should
be interpreted as a conditional and preliminary constraint rather than as a
full CMB-level determination or evidence for a resolution of the Hubble
tension.}


\begin{document}
\maketitle
\flushbottom

\section{Introduction}
The standard $\Lambda$CDM model provides a remarkably successful description
of the observed Universe accounting for the expansion history, the growth of
large-scale structure and the main features of the cosmic microwave background
(CMB) temperature and polarization anisotropies with a small number of
parameters \cite{peeb,hu,planck}. In particular, the acoustic structure of the CMB
provides a sensitive probe of the physical conditions in the early Universe
before and during recombination while supernova and baryon acoustic
oscillation (BAO) measurements constrain the subsequent expansion history.
The precision of these observables also makes them sensitive to small
departures from the standard evolution of the radiation sector.

Scalar fields that become relevant in the pre-recombination Universe provide
a well-motivated setting in which such departures may arise. Early dark energy
(EDE) models, for example, illustrate how an additional scalar degree of freedom
can affect the expansion history at early times while becoming subdominant at
later epochs \cite{kamion,bis1}. More generally, if an early-Universe scalar field
interacts with photons, energy exchange between the two sectors can modify the
standard photon dilution law and thereby alter the thermal history around
recombination. Scalar field interactions and related nonstandard
pre-recombination dynamics have therefore been studied in a variety of
cosmological contexts \cite{amen,val,clemson,salvatelli,costa,bis0}.

An especially direct probe of modified photon evolution is the CMB
temperature--redshift relation. In the standard adiabatic cosmology, the photon
energy density scales as
$\rho_\gamma\propto a^{-4}$ and the CMB temperature evolves as
$T(z)=T_0(1+z)$. 
Nonstandard temperature--redshift laws have been considered in connection with
photon--axion conversion \cite{av,mirizzi}, decaying-vacuum scenarios \cite{lima_vac,jetzer2011,jetzer2012},
scalar--photon interactions \cite{avgo,carroll,hees} and cosmic opacity \cite{avgo,bassett}. Related departures from standard radiation evolution can also arise in
interacting scalar field cosmologies in which energy is exchanged between the
scalar and radiation sectors \cite{bis1, araya, garcia, bis2}.
Such modifications are constrained
both by the observed thermal properties of the CMB and by their consequences
for recombination and the acoustic structure of the anisotropy spectra
\cite{lima2,luzzi}.

In this work we consider an early Universe scalar field non-minimally coupled
to the photon sector. Rather than introducing a deviation of the photon
dilution law phenomenologically, we begin with the underlying scalar--photon
action and derive the instantaneous quantity
\[
\epsilon(a)
=
\frac{d\ln\rho_\gamma}{d\ln a}+4
=
\frac{4\sigma}{3}\frac{d\phi}{d\ln a},
\]
which directly measures the departure of the photon energy density from the
standard $a^{-4}$ scaling. We then restrict the present analysis to the
constant-$\epsilon$ limit,
$\epsilon(a)=\epsilon$, corresponding to a controlled one-parameter
realization of the action-derived photon-sector modification. In this limit,
\[
\rho_\gamma(a)\propto a^{-4+\epsilon},
\qquad
T(z)=T_0(1+z)^{1-\epsilon/4},
\]
with the standard photon sector recovered continuously for $\epsilon=0$.
The possible dynamical evolution of $\epsilon(a)$ is not considered here.

The action-level formulation also allows the associated scalar--photon
perturbation system to be derived consistently. For theoretical completeness,
we formulate the key linear perturbation equations in synchronous gauge and
derive the corresponding regular adiabatic initial conditions for the
constant-$\epsilon$ branch. The detailed derivation is collected in
Appendix~B. These equations establish the perturbative completion associated
with the same interaction that generates the modified background photon
evolution, but they are not implemented in the numerical CLASS analysis of
the present work.

For the numerical analysis, we implement the constant-$\epsilon$ photon
density and temperature evolution in CLASS and validate the numerical
background solution directly against the analytic scaling. We then follow how
the modified photon thermal history propagates into the free-electron fraction,
the visibility function and the CMB temperature-anisotropy spectrum. These
calculations provide numerical diagnostics of the characteristic
pre-recombination signatures of the coupling. The present CLASS realization
does not yet incorporate the full coupled scalar--photon perturbation system
derived above, so the CMB spectra are not used in the likelihood analysis.

We additionally interface the modified CLASS calculation with MontePython and
perform an observational analysis using the Pantheon+SH0ES supernova likelihood
together with BAO measurements. This combination probes both the late-time
distance scale and (through the model-dependent sound horizon at the baryon
drag epoch) the altered pre-recombination photon evolution. We therefore use it
to obtain a first constraint on the action-derived constant-$\epsilon$ sector.
The resulting posterior should be interpreted within the adopted data combination, parametrization
and fixed early Universe setup; a full CMB likelihood analysis is
left for a subsequent extension.

The paper is organized as follows. Section~2 develops the scalar--photon
coupling and derives the modified photon density, temperature and background
expansion histories. Section~3 describes the background-level CLASS
implementation, validates the analytic scaling and presents the resulting
diagnostic recombination, visibility-function and CMB temperature-spectrum
responses. Section~4 presents the Pantheon+SH0ES+BAO likelihood analysis and
posterior constraints. Section~5 summarizes the main results and discusses
future extensions. Appendix~A relates the constant-$\epsilon$ scaling branch
to the scalar potential, while Appendix~B gives the complete synchronous-gauge
perturbation system and regular adiabatic initial conditions.

~~~~~~~~~~~~~~~~~~~~~~~~~~~~~~~~~~~~~~~~~~~~~~~~~~~~~~~~~~~~~~~~~~~~~~~~~~~~~~~~~~~~~~~~~~~~~~~~~~~~~~~~~~~~~~~~~
\section{EDE--photon coupling and background evolution}

\subsection{Coupling framework and temperature evolution}

The action for the system is taken to be
\footnote{Throughout this work we use reduced Planck units in which
$8\pi G=c=\hbar=1$.}
\begin{equation}
S=\int d^4x\sqrt{-g}\left\{
\frac{1}{2}R
-\frac{1}{2}g^{\mu\nu}\nabla_{\mu}\phi\nabla_{\nu}\phi
-V(\phi)
+e^{-\sigma\phi}L
\right\},
\label{a1}
\end{equation}
where $\phi$ is a minimally coupled scalar field with potential
$V(\phi)$ and $\sigma$ controls its non-minimal coupling exclusively
to the photon sector. The other cosmological components including
baryons, CDM and neutrinos remain uncoupled from
$\phi$ and obey their standard conservation equations. We use this
framework to describe an EDE-photon interaction during the
radiation-dominated epoch before recombination.

Varying the action with respect to $g_{\mu\nu}$ and $\phi$ gives
\begin{equation}
G_{\mu\nu}
=
T^{(\phi)}_{\mu\nu}
+
e^{-\sigma\phi}T_{\mu\nu},
\end{equation}
and
\begin{equation}
\nabla_{\mu}\nabla^{\mu}\phi
-
V_{,\phi}(\phi)
=
\sigma e^{-\sigma\phi}L,
\end{equation}
where
\begin{equation}
T^{(\phi)}_{\mu\nu}
=
\nabla_{\mu}\phi\nabla_{\nu}\phi
-
\frac{1}{2}g_{\mu\nu}(\nabla\phi)^2
-
g_{\mu\nu}V(\phi),
\end{equation}
and
\begin{equation}
T_{\mu\nu}
=
-\frac{2}{\sqrt{-g}}
\frac{\delta(\sqrt{-g}L)}{\delta g^{\mu\nu}}.
\end{equation}
The Bianchi identity $\nabla^{\mu}G_{\mu\nu}=0$ then implies
\begin{equation}
\nabla^{\mu}T^{(\phi)}_{\mu\nu}
=
Q_{\nu},
\end{equation}
and
\begin{equation}
\nabla^{\mu}\left(e^{-\sigma\phi}T_{\mu\nu}\right)
=
-Q_{\nu},
\end{equation}
Since radiation provides the dominant contribution to the coupled sector
during the epoch of interest, we take
${\cal L}\simeq{\cal L}_{\gamma}$. Adopting the on-shell perfect fluid
representation ${\cal L}_{\gamma}=p_{\gamma}$ \cite{schutz,brown,harko,faraoni2009} and using
$p_{\gamma}=\rho_{\gamma}/3$, we obtain
\begin{equation}
Q_{\nu}
=
\sigma e^{-\sigma\phi}{\cal L}\nabla_{\nu}\phi
\simeq
\frac{\sigma}{3}e^{-\sigma\phi}\rho_{\gamma}\nabla_{\nu}\phi .
\end{equation}

For a spatially flat FLRW spacetime, the corresponding background equations
are
\begin{equation}
\ddot{\phi}
+
3H\dot{\phi}
+
V_{,\phi}(\phi)
=
-\frac{1}{3}\sigma e^{-\sigma\phi}\rho_{\gamma},
\label{a2}
\end{equation}
and
\begin{equation}
\dot{\rho}_{\gamma}
+
4H\rho_{\gamma}
=
\frac{4}{3}\sigma\dot{\phi}\rho_{\gamma},
\label{a3}
\end{equation}
while the Friedmann and the Raychaudhuri equations during radiation domination are\footnote{The approximate equalities in equations~(11) and~(12) are intended
to isolate the background dynamics of the interacting scalar--photon sector.
Subdominant uncoupled contributions including neutrinos are therefore omitted
at this stage and are restored in the full cosmological background evolution
considered below.}
\begin{equation}
3H^2 \simeq 
e^{-\sigma\phi}\rho_{\gamma}
+
\rho_{\phi}.
\label{a4s}
\end{equation}
\begin{equation}
-2\dot H
\simeq
\frac{4}{3}e^{-\sigma\phi}\rho_{\gamma}
+\dot{\phi}^{\,2}.
\label{eq:raychaudhuri}
\end{equation}
Here
\begin{equation}
\rho_\phi
=
\frac{1}{2}\dot{\phi}^{\,2}+V(\phi),
\qquad
p_\phi
=
\frac{1}{2}\dot{\phi}^{\,2}-V(\phi).
\label{eq:scalar-rho-pressure}
\end{equation}
Equation~(\ref{a3}) integrates to
\begin{equation}
\rho_{\gamma}
\propto
a^{-4}e^{\frac{4}{3}\sigma\phi}.
\label{a5}
\end{equation}

We define the instantaneous departure from the standard photon dilution law by
\begin{equation}
\frac{d\ln\rho_{\gamma}}{d\ln a}
=
-4+\epsilon(a),
\label{a6}
\end{equation}
or equivalently,
\begin{equation}
\epsilon(a)
\equiv
\frac{d\ln\rho_{\gamma}}{d\ln a}+4
=
\frac{4\sigma}{3}\frac{d\phi}{d\ln a}.
\label{a7}
\end{equation}
Thus, $\epsilon(a)$ measures the local effect of the energy exchange on the
photon dilution rate. Positive values correspond to a slower dilution of the
photon energy density than $\rho_{\gamma}\propto a^{-4}$ whereas negative
values correspond to faster dilution. We restrict attention to small
departures $|\epsilon(a)|\ll1$.

To obtain the corresponding temperature evolution, we combine
equation~(\ref{a3}) with the blackbody relation
$\rho_{\gamma}\propto T^4$.
\footnote{We assume that the photon bath remains in local thermal equilibrium
with vanishing chemical potential \cite{lima_vac}, so that the blackbody
relation $\rho_{\gamma}\propto T^4$ remains valid in the presence of
homogeneous energy exchange with the scalar field.}
This gives
\begin{equation}
\frac{\dot{T}}{T}
+
H
=
\frac{\sigma}{3}\dot{\phi}.
\label{a8}
\end{equation}
Using equation~(\ref{a7}), the general temperature-redshift relation can be
written as
\begin{equation}
T(z)
=
T_0(1+z)
\exp\left[
-\frac{1}{4}
\int_{0}^{z}
\epsilon(\tilde z)\,
d\ln(1+\tilde z)
\right].
\label{eq:T-general}
\end{equation}
When $\epsilon(a)$ is approximately constant over the epoch of interest, this
relation reduces to
\begin{equation}
T(z)
=
T_0(1+z)^{1-\epsilon/4}.
\label{a14}
\end{equation}

This relation provides the basis for the subsequent analysis. In the
underlying scalar field model, $\epsilon(a)$ is determined by the evolution of
$\phi$ through equation~(\ref{a7}). Here, however, we restrict attention to its
constant limit. The standard photon temperature-redshift relation is recovered
for $\epsilon=0$.

\subsection{Background evolution in the constant-$\epsilon$ limit}

The constant-$\epsilon$ limit is of particular interest because it corresponds
to a scaling branch in which the logarithmic deviation of the photon dilution
rate from its standard adiabatic value remains constant. From
equation~(\ref{a7}), $\epsilon=\mathrm{const.}$ implies
$d\phi/d\ln a=3\epsilon/(4\sigma)=\mathrm{const.}$, so that the homogeneous
scalar field evolves logarithmically with the scale factor. Thus, the
constant-$\epsilon$ condition restricts the system to a well-defined
background trajectory of the underlying scalar--photon dynamics rather than
introducing an independent phenomenological modification of the photon
dilution law. The relation of this constant-$\epsilon$ scaling branch to the
corresponding scalar-field potential is discussed in
Appendix~A.\\
For constant $\epsilon$, the photon energy density evolves as
\begin{equation}
\rho_{\gamma}(a)
=
\rho_{\gamma0}a^{-4+\epsilon},
\label{b3}
\end{equation}
or equivalently,
\begin{equation}
\rho_{\gamma}(z)
=
\rho_{\gamma0}(1+z)^{4-\epsilon},
\label{eq:rho-gamma-z}
\end{equation}
where $\rho_{\gamma0}$ is the present photon energy density. Using
$d\phi/d\ln a=3\epsilon/(4\sigma)$, the photon contribution that appears in
the Einstein equations evolves as
\begin{equation*}
e^{-\sigma\phi}\rho_{\gamma}(z)
=
e^{-\sigma\phi_0}\rho_{\gamma0}(1+z)^{4-\epsilon/4}.
\end{equation*}
On the corresponding scaling branch, the scalar energy density obeys
$\rho_{\phi}(z)=\rho_{\phi0}(1+z)^{4-\epsilon/4}$.
Normalizing the standard and extended models to the same present-day photon density gives
\begin{equation}
\frac{\rho_{\gamma}(z)}
{\rho_{\gamma}^{\Lambda\mathrm{CDM}}(z)}
=
(1+z)^{-\epsilon}.
\label{b4}
\end{equation}
Similarly, equation~(\ref{a14}) gives
\begin{equation}
\frac{T(z)}{T_0(1+z)}
=
(1+z)^{-\epsilon/4}.
\label{b7}
\end{equation}
At fixed present-day normalization, negative values of $\epsilon$ increase the
photon temperature and energy density at high redshift relative to
$\Lambda$CDM whereas positive values reduce them.

To determine the corresponding effect on the background expansion, we retain
the standard matter, neutrino and cosmological constant components, while the
interacting scalar--photon sector contributes through
$e^{-\sigma\phi}\rho_{\gamma}+\rho_{\phi}$. For a spatially flat reference
cosmology,
\begin{equation}
E_{\Lambda\mathrm{CDM}}^2(z)
=
\Omega_m(1+z)^3
+
\Omega_{\gamma}(1+z)^4
+
\Omega_{\nu}(1+z)^4
+
\Omega_{\Lambda},
\label{eq:E-lcdm}
\end{equation}
where $E(z)\equiv H(z)/H_0$. We define
$\Omega_{\phi}\equiv\rho_{\phi0}/(3H_0^2)$.
To compare the extended model with the same spatially flat reference
cosmology and the same present-day values of $H_0$, $\Omega_m$,
$\Omega_{\nu}$ and $\Omega_{\Lambda}$, we impose the matched
present-day normalization
$e^{-\sigma\phi_0}\Omega_{\gamma}+\Omega_{\phi}=\Omega_{\gamma}$.
The corresponding expression in the constant-$\epsilon$ model is
\begin{equation}
E_{\epsilon}^2(z)
=
\Omega_m(1+z)^3
+
\Omega_{\nu}(1+z)^4
+
\Omega_{\Lambda}
+
\Omega_{\gamma}(1+z)^{4-\epsilon/4}.
\label{eq:E-epsilon}
\end{equation}
The neutrino contribution remains standard and uncoupled, while the
interacting scalar--photon sector is represented by the final scaling term.
The fractional change in the Hubble rate is therefore
\begin{equation}
\frac{\Delta H}{H^{\Lambda\mathrm{CDM}}}
\equiv
\frac{H_{\epsilon}(z)-H^{\Lambda\mathrm{CDM}}(z)}
{H^{\Lambda\mathrm{CDM}}(z)}
=
\left[
\frac{E_{\epsilon}^2(z)}
{E_{\Lambda\mathrm{CDM}}^2(z)}
\right]^{1/2}
-1.
\label{eq:deltaH-analytic}
\end{equation}
These expressions provide the analytic reference for the numerical
implementation discussed in the following section.

For theoretical completeness, the coupled linear perturbation equations and regular adiabatic initial conditions associated with the same scalar--photon interaction are collected in Appendix~B; they are not evolved in the CLASS calculations below, which remain restricted to the homogeneous constant-$\epsilon$ background.

~~~~~~~~~~~~~~~~~~~~~~~~~~~~~~~~~~~~~~~~~~~~~~~~~~~~~~~~~~~~~~~~~~~~~~~~~~~~~~~~~~~~~~~~~~~~~~~~~~~~~~~
\section{Numerical results}
We implement the constant-$\epsilon$ photon evolution in the Boltzmann code
CLASS \cite{class}. The present implementation is deliberately restricted to
the homogeneous photon sector: the photon energy density and the
temperature--redshift relation are modified according to
equations~(\ref{b3}) and~(\ref{a14}), respectively, while matter, neutrinos
and the cosmological constant retain their standard background evolution.
The coupled scalar--photon perturbation equations and regular adiabatic
initial conditions collected in Appendix~B are not evolved in these runs. We first validate the numerical background evolution
against the analytic constant-$\epsilon$ prediction and then use the same
background implementation to examine the diagnostic response of the
recombination history, visibility function and CMB temperature spectrum.

\subsection{CLASS implementation and background validation}
We evaluate the model for
$\epsilon=-0.04,-0.02,0,+0.02,$ and $+0.04$. For this
background-validation test, the primordial helium fraction is kept fixed.
The photon density ratio obtained from CLASS is compared with the analytic
prediction
given in equation~(\ref{b4}). A direct comparison at $z=1100$ shows that the numerical photon density
ratios agree with the analytic prediction of equation~(\ref{b4}) within numerical
precision confirming the implementation of the modified photon background
evolution.

\par\medskip
\begin{center}
\safeincludegraphics[width=0.70\textwidth]{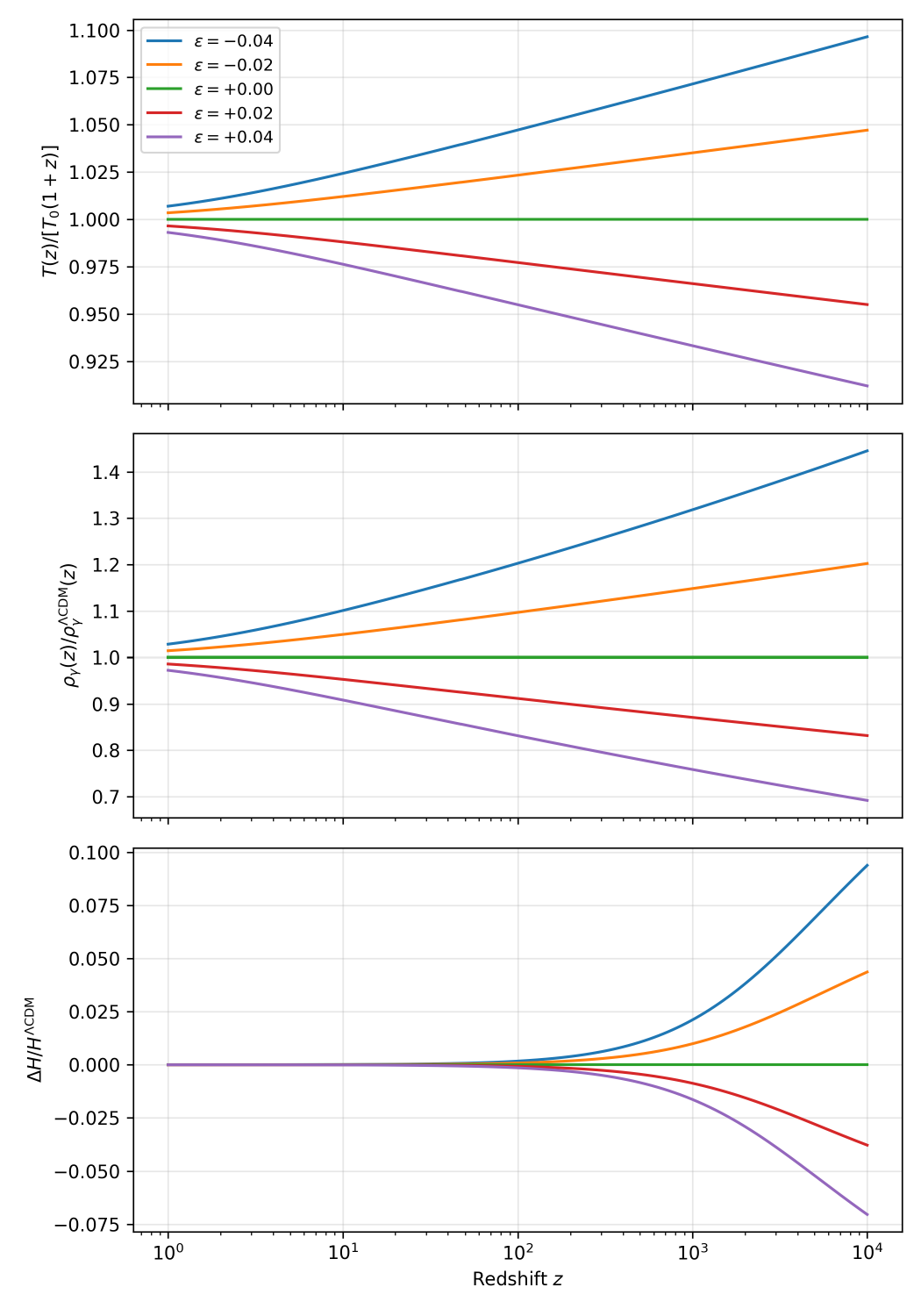}
\end{center}
\refstepcounter{figure}
\label{fig:class-background}
\noindent{\small\textbf{Figure~\thefigure:}
Background response of the constant-$\epsilon$ model. The upper panel shows
the temperature ratio $T(z)/[T_0(1+z)]$, the middle panel shows the
photon density ratio
$\rho_{\gamma}(z)/\rho_{\gamma}^{\Lambda\mathrm{CDM}}(z)$ obtained from the
modified CLASS background output and the lower panel shows the fractional
Hubble rate deviation defined in equation~(\ref{eq:deltaH-analytic}). The
standard photon evolution is recovered for $\epsilon=0$.
\par}
\par\medskip

Figure~\ref{fig:class-background} shows the corresponding changes in the
photon temperature, photon energy density and expansion rate. At fixed
present-day normalization, negative values of $\epsilon$ increase the photon
temperature and energy density at high redshift and therefore produce a
larger Hubble rate relative to $\Lambda\mathrm{CDM}$. Positive values of
$\epsilon$ produce the opposite behaviour. The agreement between the
analytic and numerical photon density evolution validates the background
implementation.

\subsection{Recombination history and visibility function}
\label{subsec:recombination}

The modified temperature-redshift relation changes the thermal conditions
during recombination. We therefore examine the free-electron fraction
\begin{equation}
x_e(z)\equiv\frac{n_e(z)}{n_{\rm H}(z)},
\label{eq:xe_definition}
\end{equation}
where $n_e$ is the number density of free electrons and $n_{\rm H}$ is the
number density of hydrogen nuclei. The free-electron fraction describes the
ionization state of the primordial plasma and controls the transition from
the tightly coupled photon-baryon fluid to a transparent Universe
\cite{peeb,hu,planck}.

The corresponding last-scattering distribution is characterized by the
visibility function shown here as a function of redshift $z$. The free-electron fraction and
visibility function obtained from the modified CLASS calculation are shown
in Figure~\ref{fig:recombination-diagnostics}.

\par\medskip
\begin{center}
\safeincludegraphics[width=0.48\textwidth]
{recombination_xe_eps_pm002.pdf}
\hfill
\safeincludegraphics[width=0.48\textwidth]
{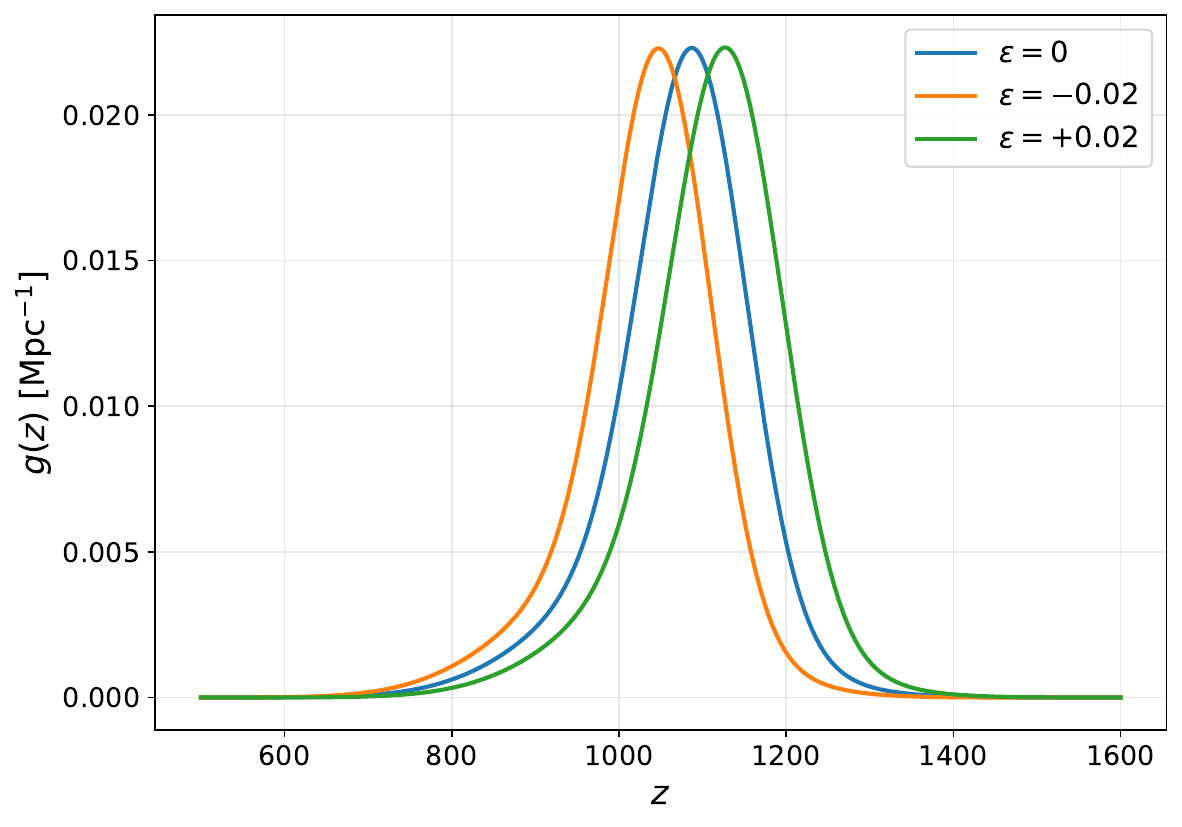}
\end{center}
\refstepcounter{figure}
\label{fig:recombination-diagnostics}
\noindent{\small\textbf{Figure~\thefigure:}
Recombination diagnostics for representative values of $\epsilon$. The left panel shows the free-electron fraction $x_e(z)$ while the right
panel shows the visibility function plotted as a function of
redshift $z$. Negative values of $\epsilon$ delay
recombination and shift the last-scattering surface toward lower redshift
whereas positive values produce the opposite shifts.
\par}
\par\medskip

At fixed redshift, negative values of $\epsilon$ correspond to a higher
photon temperature than in the standard case. Recombination is therefore delayed, shifting both the recombination
transition in $x_e(z)$ and the visibility peak toward lower redshift. Positive values of $\epsilon$ lower the photon
temperature, causing recombination and last scattering to occur at higher
redshift. The overall shape of the visibility function remains qualitatively
similar to that of the standard model.

\subsection{CMB temperature spectrum response}
\label{subsec:cmb-response}

The modified photon density and recombination history also affect the CMB
temperature anisotropy spectrum. We examine the temperature power spectrum
\[
D_{\ell}^{TT}\equiv
\frac{\ell(\ell+1)}{2\pi}C_{\ell}^{TT},
\]
together with its fractional difference relative to the standard case,
\begin{equation}
\frac{D_{\ell}^{TT}(\epsilon)}
     {D_{\ell}^{TT}(0)}-1.
\label{eq:tt-fractional}
\end{equation}

\par\medskip
\begin{center}
\safeincludegraphics[width=0.48\textwidth]
{cmb_TT_spectrum_eps_pm002.pdf}
\hfill
\safeincludegraphics[width=0.48\textwidth]
{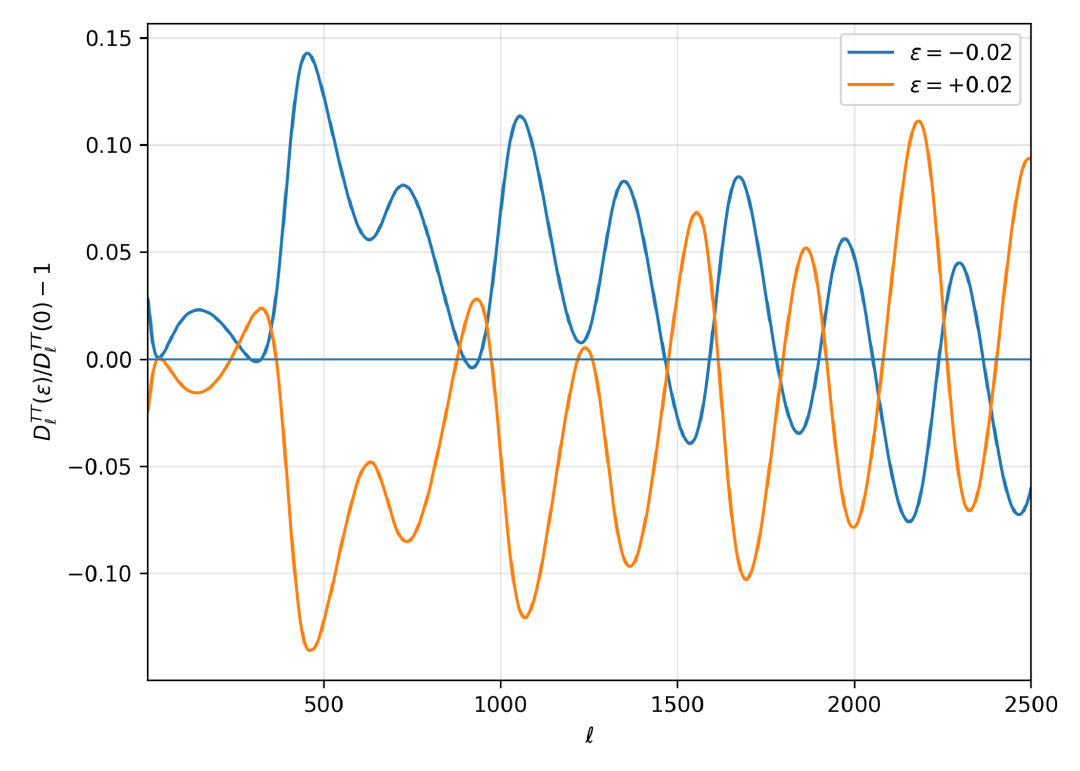}
\end{center}
\refstepcounter{figure}
\label{fig:cmb-diagnostics}
\noindent{\small\textbf{Figure~\thefigure:}
Diagnostic response of the CMB temperature spectrum to the
constant-$\epsilon$ background modification. The left panel shows
$D_{\ell}^{TT}$ for $\epsilon=0$ and representative non-zero values of
$\epsilon$. The right panel shows the corresponding fractional difference
$D_{\ell}^{TT}(\epsilon)/D_{\ell}^{TT}(0)-1$. The coupled scalar--photon
perturbation system is not evolved in these calculations; the curves are
therefore diagnostic responses to the modified background and should not be
interpreted as fully self-consistent predictions of the interacting model.
\par}
\par\medskip

As shown in Figure~\ref{fig:cmb-diagnostics}, the modified photon evolution
changes both the relative amplitudes and the positions of the acoustic peaks.
The oscillatory structure of the fractional response reflects changes in
both the amplitudes and phases of the acoustic oscillations. These effects
arise from the combined modification of the background photon density, the
temperature-redshift relation and the recombination history.

The results presented in this section illustrate how the modified
constant-$\epsilon$ background photon evolution affects recombination and
produces corresponding diagnostic changes in the CMB temperature spectrum.
Because the coupled scalar--photon perturbation system is not evolved, these
spectra are not full predictions of the interacting model. The recombination
and CMB outputs are used only as diagnostic responses and do not enter the
likelihood analysis. A complete CMB constraint would require a self-consistent
implementation of the coupled scalar--photon perturbation equations together
with CMB temperature, polarization and lensing likelihoods.

~~~~~~~~~~~~~~~~~~~~~~~~~~~~~~~~~~~~~~~~~~~~~~~~~~~~~~~~~~~~~~~~~~~~~~~~~~~~~~~~~~~~~~~~~~~~~~~~~~~~~~~~~~~~~~~
\section{Data, methodology and posterior constraints}

In this section we use the background-level CLASS implementation described in
Section~3 to obtain preliminary posterior constraints on the constant-$\epsilon$
photon sector deviation parameter. The analysis is carried out with
MontePython \cite{montepython,montepython3} interfaced with the modified version
of CLASS.

\subsection{Datasets}

The observational data combination consists of the Pantheon+SH0ES supernova likelihood together with the BAO likelihoods adopted in the MontePython analysis. Although the BAO measurements are obtained at low and intermediate redshifts, their theoretical predictions depend on the drag sound horizon $r_d$, evaluated at the baryon drag epoch. The resulting constraint on $\epsilon$ therefore reflects both the late time distance–redshift relation and the model-dependent modification of the pre-recombination photon evolution and drag scale.


\subsection{Parameter and sampling setup}

We consider two models. In the extended model the sampled parameters are
\[
\left\{\Omega_m,\ H_0,\ M,\ \epsilon\right\},
\]
where $M$ is the supernova absolute-magnitude nuisance parameter. In the
baseline model the sampled parameters are
\[
\left\{\Omega_m,\ H_0,\ M\right\},
\]
with $\epsilon=0$ fixed. Spatial flatness is imposed and $\Omega_\Lambda$ is
treated as a derived parameter. Parameters not included in these sets
including the physical baryon density and neutrino sector quantities required
by CLASS are held fixed at the fiducial values of the adopted MontePython
configuration. The primordial helium mass fraction $Y_{\rm He}$ is also fixed
and no BBN consistency relation is imposed. These choices make the result a
conditional, preliminary constraint rather than a fully BBN or CMB-consistent
parameter inference.

\subsection{Posterior constraints}

For the extended constant-$\epsilon$ model, the marginalized posterior gives
\begin{equation}
\epsilon=0.0230\pm0.0065,
\end{equation}
with the corresponding $16$th--$84$th percentile credible interval
\begin{equation}
0.0168<\epsilon<0.0293.
\end{equation}
Thus, within the adopted Pantheon+SH0ES+BAO likelihood and fixed-parameter
setup, the posterior is shifted toward positive values of $\epsilon$.

The remaining marginalized constraints are summarized in
Table~\ref{tab:posterior_constraints}. The entries are posterior means and
standard deviations after discarding the first $50\%$ of each chain as burn-in.

\begin{table}[!htbp]
\centering
\begin{tabular}{lcccc}
\hline
Model & $\Omega_m$ & $H_0\,[{\rm km\,s^{-1}\,Mpc^{-1}}]$ & $M$ & $\epsilon$ \\
\hline
$\epsilon$ free
&
$0.3137\pm0.0125$
&
$73.61\pm0.94$
&
$-19.252\pm0.028$
&
$0.0230\pm0.0065$
\\
$\epsilon=0$
&
$0.3314\pm0.0124$
&
$72.77\pm0.97$
&
$-19.270\pm0.029$
&
$0\ {\rm fixed}$
\\
\hline
\end{tabular}
\caption{Preliminary posterior constraints from the adopted
Pantheon+SH0ES+BAO likelihood with fixed early-Universe parameters,
including $Y_{\rm He}$. The entries are
posterior means and standard deviations after a $50\%$ burn-in. The baseline
model has $\epsilon=0$ fixed whereas the extended model samples $\epsilon$ as
an additional cosmological parameter.
\label{tab:posterior_constraints}}
\end{table}

The corresponding one- and two-dimensional marginalized posterior
distributions are shown in Figure~\ref{fig:triangle_epsilon}. The strong
$M$--$H_0$ correlation is expected for a supernova-calibrated distance analysis.
The figure also shows correlations between $\epsilon$ and the sampled
background parameters so that its marginalized constraint remains
conditional on the adopted parameterization and on the fixed early Universe
quantities.

\par\medskip
\begin{center}
\safeincludegraphics[width=0.95\textwidth]{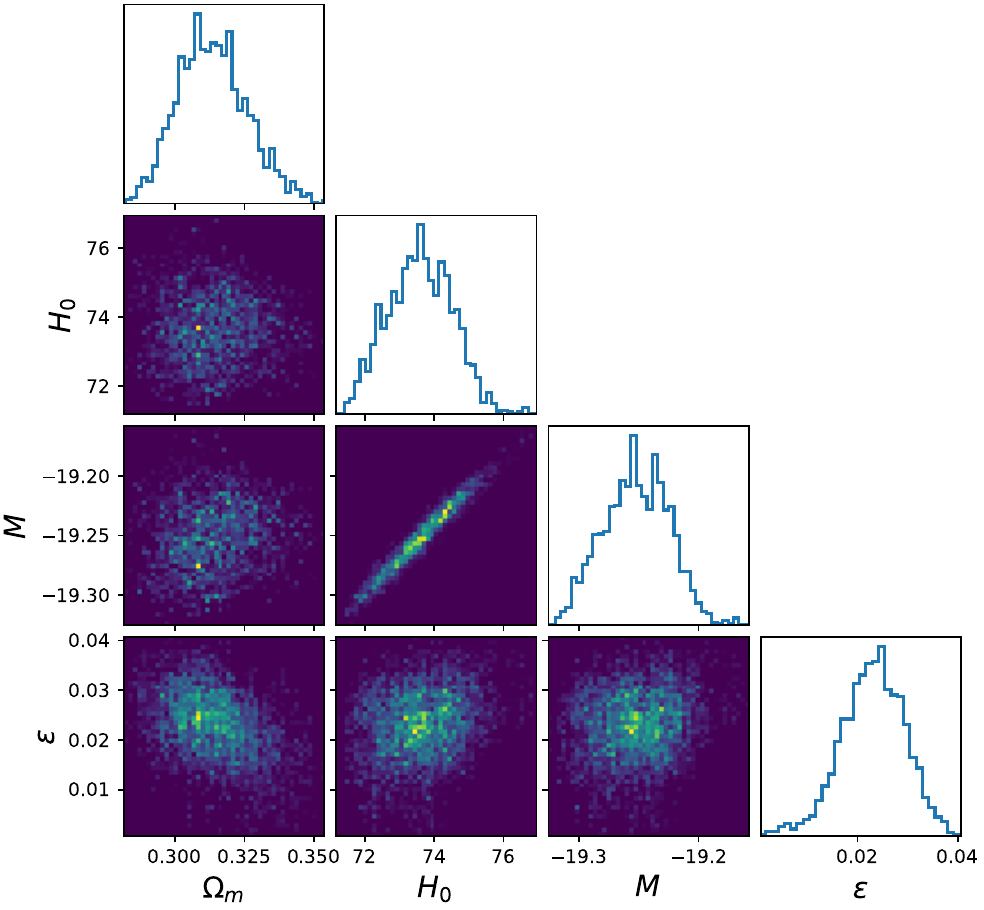}
\end{center}
\refstepcounter{figure}
\label{fig:triangle_epsilon}
\noindent{\small\textbf{Figure~\thefigure:} Posterior distributions and
parameter degeneracies for the extended constant-$\epsilon$ model using the
adopted Pantheon+SH0ES+BAO likelihood. The sampled parameters are the matter
density parameter $\Omega_m$, the Hubble constant $H_0$, the supernova
absolute-magnitude nuisance parameter $M$ and the photon sector deviation
parameter $\epsilon$.\par}
\par\medskip

The preference for positive $\epsilon$ should be interpreted cautiously.
In particular, BAO observables retain sensitivity to the model-dependent
drag scale while the physical baryon density, neutrino sector and
$Y_{\rm He}$ are fixed in the present setup.

~~~~~~~~~~~~~~~~~~~~~~~~~~~~~~~~~~~~~~~~~~~~~~~~~~~~~~~~~~~~~~~~~~~~~~~~~~~~~~~~~~~~~~~~~~~~~~~~

\section{Conclusions}
In this work, we investigated a modification of the photon sector motivated
by an early-time interaction between a scalar field and radiation. Starting
from a scalar field action non-minimally coupled to the photon sector Lagrangian, we
introduced the instantaneous deviation parameter $\epsilon(a)$ which measures
the departure of the photon energy density from the standard adiabatic
scaling. We then focused on the constant-$\epsilon$ limit in which the
modification is characterized by a single action-derived parameter controlling
both the photon-density evolution and the CMB temperature--redshift relation.

We implemented the constant-$\epsilon$ photon evolution in CLASS and verified
that the numerical background output reproduces the analytic scaling for the
representative values considered. At fixed present-day normalization, negative
values of $\epsilon$ increase the photon temperature and energy density at high
redshift and consequently increase the expansion rate relative to the standard
case. Positive values of $\epsilon$ produce the opposite behaviour. This
agreement validates the numerical implementation of the modified photon
evolution at the homogeneous background level.

We also examined the diagnostic response of the recombination history,
visibility function and CMB temperature spectrum. The modified
temperature--redshift relation changes the thermal conditions during
recombination, shifting the free-electron fraction and the last-scattering
surface. It also produces scale-dependent changes in the amplitudes and
positions of the acoustic peaks. These results illustrate how the modified
constant-$\epsilon$ background photon evolution affects recombination and
produces corresponding diagnostic changes in CMB observables. The calculated
spectra are therefore used only as diagnostic responses to the background
modification and are not interpreted as perturbatively self-consistent
predictions of the interacting model, since the full coupled scalar--photon
perturbation system has not yet been implemented and no CMB likelihood
analysis has been performed in the present work.

As a preliminary statistical application, we combined the modified CLASS
implementation with MontePython and used the Pantheon+SH0ES supernova
likelihood together with BAO measurements. The resulting marginalized
constraint is
\[
\epsilon=0.0230\pm0.0065,
\]
with the posterior centered on positive values within the adopted likelihood
combination, priors and fixed-parameter setup. In the convention used here,
positive $\epsilon$ corresponds to a lower photon temperature and photon
energy density at high redshift relative to the standard adiabatic evolution.
Because the BAO predictions depend on the model-dependent drag scale, and
because several early Universe quantities, including the physical baryon
density, neutrino-sector parameters and $Y_{\rm He}$, are held fixed, this
result should be regarded as a conditional and preliminary constraint.

The constant-$\epsilon$ model provides a minimal first step toward a dynamical
description of scalar--radiation energy exchange. A natural extension is to
replace the constant parameter by a redshift-dependent quantity
$\epsilon(z)$ determined dynamically by the evolution of the coupled scalar
field. Such a formulation would allow the photon-sector deviation to evolve
consistently with the scalar potential, field trajectory and time-dependent
energy-transfer rate. Future measurements of the CMB temperature--redshift
relation, CMB anisotropies, BAO observables and calibrated supernova distances
will provide increasingly sensitive tests of constant and evolving deviations
from standard photon evolution
\cite{cmbs4,simons,desi2024,euclid}. A dynamical and perturbatively
self-consistent extension is left for future work.\\\\
~~~~~~~~~~~~~~~~~~~~~~~~~~~~~~~~~~~~~~~~~~~~~~~~~~~~~~~~~~~~~~~~~~~~~~~~~~~~~~~~~~~~~~~~~~~~~~~~
\setcounter{equation}{0}
\renewcommand{\theequation}{A\arabic{equation}}
\textbf{Appendix A: Constant-$\epsilon$ scaling solution and the form of the scalar potential}\\
\label{app:constant_epsilon_scaling}

In this appendix we clarify the relation between the constant-$\epsilon$
limit adopted in the main text and the form of the scalar field potential.
Scaling solutions of canonical scalar fields are well known to be closely
associated with exponential potentials. In the uncoupled case, exponential
potentials admit solutions in which the scalar-field energy density tracks
the dominant barotropic component \cite{copeland,ferreira}, while analogous
scaling solutions also arise in coupled scalar--fluid systems
\cite{amen, amendola_scale}. Here we derive the corresponding relation for
the scalar--photon interaction considered in this work.

It is convenient to introduce the number of e-folds,
\begin{equation}
N\equiv\ln a.
\label{eq:app_N}
\end{equation}
From equation~(\ref{a7}), the deviation parameter is
\begin{equation}
\epsilon(a)
=
\frac{4\sigma}{3}\frac{d\phi}{dN}.
\label{eq:app_epsilon}
\end{equation}
Consequently, on the constant-$\epsilon$ branch,
\begin{equation}
\frac{d\phi}{dN}
=
\frac{3\epsilon}{4\sigma}
\equiv q
=
\mathrm{const.}
\label{eq:app_phiN}
\end{equation}
The scalar field therefore evolves linearly with $N$ or equivalently
logarithmically with the scale factor,
\begin{equation}
\phi(a)
=
\phi_{\star}
+
\frac{3\epsilon}{4\sigma}
\ln\!\left(\frac{a}{a_{\star}}\right),
\label{eq:app_phi_a}
\end{equation}
where $a_{\star}$ and $\phi_{\star}$ denote arbitrary reference values.

The photon density obtained from equation~(\ref{a5}) is
\begin{equation}
\rho_{\gamma}
=
\rho_{\gamma\star}
\left(\frac{a}{a_{\star}}\right)^{-4}
\exp\!\left[
\frac{4\sigma}{3}
(\phi-\phi_{\star})
\right].
\label{eq:app_rhog_general}
\end{equation}
Using equation~\eqref{eq:app_phi_a}, this becomes
\begin{equation}
\rho_{\gamma}
=
\rho_{\gamma\star}
\left(\frac{a}{a_{\star}}\right)^{-4+\epsilon},
\label{eq:app_rhog_scaling}
\end{equation}
which is the constant-$\epsilon$ photon dilution law used in the main text.

Because the photon sector appears in the Einstein equations multiplied by
the coupling factor $e^{-\sigma\phi}$, the quantity relevant for the
background gravitational dynamics is the explicit combination
$e^{-\sigma\phi}\rho_{\gamma}$. Using equation~\eqref{eq:app_rhog_general},
this contribution can be written as
\begin{equation}
e^{-\sigma\phi}\rho_{\gamma}
=
e^{-\sigma\phi_{\star}}\rho_{\gamma\star}
\left(\frac{a}{a_{\star}}\right)^{-4}
\exp\!\left[
\frac{\sigma}{3}(\phi-\phi_{\star})
\right].
\label{eq:app_rhocoupled_general}
\end{equation}
Equation~\eqref{eq:app_phi_a} gives
\begin{equation}
\exp\!\left[
\frac{\sigma}{3}(\phi-\phi_{\star})
\right]
=
\left(\frac{a}{a_{\star}}\right)^{\epsilon/4},
\label{eq:app_coupling_scaling}
\end{equation}
and hence
\begin{equation}
e^{-\sigma\phi}\rho_{\gamma}
=
e^{-\sigma\phi_{\star}}\rho_{\gamma\star}
\left(\frac{a}{a_{\star}}\right)^{-4+\epsilon/4}.
\label{eq:app_rhocoupled_scaling}
\end{equation}
Thus, the constant-$\epsilon$ branch is characterized by a fixed
power-law evolution of the coupled radiation contribution.

Restricting further to the scaling fixed point for which the ratio
$\rho_\phi/(e^{-\sigma\phi}\rho_\gamma)$ remains constant, the scalar
energy density must scale with the same power of $a$ as the coupled
radiation contribution, 
\begin{equation}
\rho_{\phi}
\propto
e^{-\sigma\phi}\rho_{\gamma}
\propto
a^{-4+\epsilon/4}.
\label{eq:app_scaling_condition}
\end{equation}
This is the coupled analogue of the familiar scalar fluid scaling
solutions discussed in
Refs.~\cite{copeland,ferreira,amendola_scale}.

The same conclusion can be expressed in terms of the kinetic and potential
parts of the scalar energy density. From
equation~\eqref{eq:app_phiN},
\begin{equation}
\dot{\phi}
=
H\frac{d\phi}{dN}
=
\frac{3\epsilon}{4\sigma}H,
\label{eq:app_phidot}
\end{equation}
so that
\begin{equation}
\frac{1}{2}\dot{\phi}^{\,2}
=
\frac{1}{2}
\left(\frac{3\epsilon}{4\sigma}\right)^2 H^2.
\label{eq:app_kinetic}
\end{equation}
On the scaling branch, $H^2$ has the same fixed power-law dependence as
the total energy density,
\begin{equation}
H^2\propto a^{-4+\epsilon/4}.
\label{eq:app_H_scaling}
\end{equation}
The kinetic contribution therefore scales as
$a^{-4+\epsilon/4}$. Since
\begin{equation}
\rho_{\phi}
=
\frac{1}{2}\dot{\phi}^{\,2}
+
V(\phi),
\end{equation}
consistency of the scaling solution requires the potential contribution
to obey the same scaling law,
\begin{equation}
V(a)
=
V_{\star}
\left(\frac{a}{a_{\star}}\right)^{-4+\epsilon/4}.
\label{eq:app_V_a}
\end{equation}

Equation~\eqref{eq:app_phi_a} can be inverted to give
\begin{equation}
\ln\!\left(\frac{a}{a_{\star}}\right)
=
\frac{4\sigma}{3\epsilon}
(\phi-\phi_{\star}).
\label{eq:app_a_phi}
\end{equation}
Substituting equation~\eqref{eq:app_a_phi} into
equation~\eqref{eq:app_V_a} yields
\begin{align}
V(\phi)
&=
V_{\star}
\exp\!\left[
-\left(4-\frac{\epsilon}{4}\right)
\frac{4\sigma}{3\epsilon}
(\phi-\phi_{\star})
\right]
\nonumber\\[1mm]
&=
V_{\star}
\exp\!\left[
-\lambda(\phi-\phi_{\star})
\right],
\label{eq:app_V_exp}
\end{align}
where
\begin{equation}
\lambda=\frac{\sigma(16-\epsilon)}{3\epsilon}.
\label{eq:app_lambda}
\end{equation}

The appearance of the exponential potential can therefore be understood
directly from the structure of the constant-$\epsilon$ scaling branch.
Constancy of $\epsilon$ implies
\begin{equation}
\phi\propto\ln a,
\end{equation}
whereas the scaling solution requires the potential energy to evolve as a
fixed power of the scale factor,
\begin{equation}
V\propto a^{-4+\epsilon/4}.
\end{equation}
A power law in $a$ necessarily becomes an exponential when expressed as a
function of a field that is linear in $\ln a$. In schematic form,
\begin{equation}
\phi\propto\ln a,
\qquad
V\propto a^{-m}
\quad\Longrightarrow\quad
V(\phi)\propto e^{-\lambda\phi}.
\end{equation}
For the scalar--photon coupling considered here, the exponent is fixed
specifically by equation~\eqref{eq:app_lambda}.

This result is consistent with the established role of exponential potentials
in cosmological scaling solutions and with their
extension to systems containing a constant coupling between a scalar field
and a cosmological fluid. The relation~\eqref{eq:app_lambda},
however, is specific to the coupling and definition of $\epsilon$ adopted
in the present model. It also shows that once a constant value of
$\epsilon$ and a coupling strength $\sigma$ are specified on this branch,
the logarithmic slope of the corresponding exponential potential is no
longer an independent parameter.\\\\
\setcounter{equation}{0}
\renewcommand{\theequation}{B\arabic{equation}}
\noindent\textbf{Appendix B: Coupled perturbation system and regular adiabatic}\\
\textbf{initial conditions}\\
This appendix collects the detailed synchronous gauge perturbation
formulation associated with the same scalar--photon interaction that produces
the constant-$\epsilon$ background evolution. It provides the perturbative
completion of the model and the regular adiabatic initial conditions but the
equations collected here are not evolved in the CLASS calculations of the
present work. 

\medskip
\noindent\textbf{Perturbation definitions and closure.}\par

Following the perturbative treatment of coupled scalar fluid models
and the standard synchronous gauge formalism
\cite{liu2024,ma_bert,malik_wands}, we work in synchronous gauge,
\begin{equation}
ds^{2}
=
a^{2}(\tau)
\left[
-d\tau^{2}
+
\left(\delta_{ij}+h_{ij}\right)dx^{i}dx^{j}
\right],
\label{eq:synchronous-metric}
\end{equation}
where $\tau$ is conformal time,
$\mathcal{H}\equiv a'/a$ and a prime denotes $d/d\tau$. 
For scalar perturbations, the metric perturbation is decomposed as
\begin{equation}
h_{ij}
=
\hat{k}_{i}\hat{k}_{j}h
+
6\eta
\left(
\hat{k}_{i}\hat{k}_{j}
-
\frac{1}{3}\delta_{ij}
\right),
\label{eq:metric-decomposition}
\end{equation}
where $h$ and $\eta$ are the synchronous gauge scalar metric
perturbations. 
The scalar field
and photon density are decomposed as
\begin{equation}
\phi=\bar{\phi}+\delta\phi,
\qquad
\rho_{\gamma}
=
\bar{\rho}_{\gamma}
\left(1+\delta_{\gamma}\right),
\label{eq:perturbation-decomposition}
\end{equation}
where $\delta_{\gamma}\equiv
\delta\rho_{\gamma}/\bar{\rho}_{\gamma}$. We denote the photon velocity
divergence by
$\theta_{\gamma}\equiv ik_jv_{\gamma}^{j}$ and the photon anisotropic stress by
$\Pi_{\gamma}\equiv F_{\gamma2}/2$. The latter corresponds to the
usual photon shear variable in the synchronous gauge hierarchy
\cite{ma_bert,class}.

Perturbing the coupled scalar field equation gives
\begin{align}
\delta\phi''
+
2\mathcal{H}\delta\phi'
+
\frac{1}{2}h'\bar{\phi}'
+
\left[
k^{2}
+
a^{2}V_{,\phi\phi}
-
\frac{a^{2}\sigma^{2}}{3}
e^{-\sigma\bar{\phi}}\bar{\rho}_{\gamma}
\right]\delta\phi
=
-\frac{a^{2}\sigma}{3}
e^{-\sigma\bar{\phi}}\bar{\rho}_{\gamma}\delta_{\gamma}.
\label{eq:scalar-perturbation}
\end{align}
This equation reduces to the standard minimally coupled scalar field
perturbation equation when the interaction is removed.
The scalar field density, pressure and momentum perturbations are
\cite{liu2024,malik_wands}
\begin{subequations}
\label{eq:scalar-fluid-perturbations}
\begin{align}
\delta\rho_{\phi}
&=
\frac{\bar{\phi}'\delta\phi'}{a^{2}}
+
V_{,\phi}\delta\phi,
\\
\delta p_{\phi}
&=
\frac{\bar{\phi}'\delta\phi'}{a^{2}}
-
V_{,\phi}\delta\phi,
\\
\left(
\bar{\rho}_{\phi}+\bar{p}_{\phi}
\right)\theta_{\phi}
&=
\frac{k^{2}\bar{\phi}'\delta\phi}{a^{2}}.
\end{align}
\end{subequations}

Perturbing the covariant energy transfer equation gives the modified photon
continuity equation
\begin{equation}
\delta_{\gamma}'
+
\frac{4}{3}\theta_{\gamma}
+
\frac{2}{3}h'
=
\frac{4\sigma}{3}\delta\phi'.
\label{eq:photon-continuity-coupled}
\end{equation}
For the constant-$\epsilon$ implementation we adopt an isotropic
energy transfer closure in photon momentum space. Under this phenomenological
closure, the interaction modifies the photon monopole directly while the
photon Euler equation retains its standard synchronous gauge form
\cite{ma_bert,class},
\begin{equation}
\theta_{\gamma}'
=
k^{2}
\left(
\frac{\delta_{\gamma}}{4}
-
\Pi_{\gamma}
\right)
+
a n_e\sigma_{\rm T}
\left(
\theta_b-\theta_{\gamma}
\right).
\label{eq:photon-euler-coupled}
\end{equation}
Under the same closure, the higher photon temperature multipoles and the
polarization hierarchy retain their standard CLASS evolution equations \cite{ma_bert,class}.
The new monopole source in equation~(\ref{eq:photon-continuity-coupled})
propagates to the higher multipoles through free streaming and Thomson
scattering. Baryons, CDM and neutrinos remain uncoupled from
$\phi$ and obey their standard perturbation equations.

The background relation in equation~\eqref{a7} becomes
\begin{equation}
\epsilon
=
\frac{4\sigma}{3}
\frac{\bar{\phi}'}{\mathcal{H}},
\qquad
\bar{\phi}'
=
\frac{3\epsilon}{4\sigma}\mathcal{H}
\quad
\text{for constant }\epsilon.
\label{eq:epsilon-conformal}
\end{equation}
This condition constrains the homogeneous trajectory and does not imply
$\delta\phi=0$. The uncoupled limit must be taken in the first relation in
equation~\eqref{eq:epsilon-conformal} before solving it for $\bar{\phi}'$ since the
second relation is valid only for $\sigma\neq0$. 
In the uncoupled limit $\sigma=0$, the interaction terms vanish and
$\epsilon=0$. Equations~\eqref{eq:scalar-perturbation},
\eqref{eq:photon-continuity-coupled} and
\eqref{eq:photon-euler-coupled} then reduce to the standard minimally coupled
scalar field and photon perturbation equations.

\medskip
\noindent\textbf{Regular adiabatic initial conditions.}\par

We impose initial conditions deep in the radiation dominated epoch
following the regular synchronous gauge treatment
\cite{liu2024,ma_bert}, when
\begin{equation}
k\tau_i\ll1,
\qquad
a n_e\sigma_{\rm T}\gg\mathcal{H},
\label{eq:initial-regime}
\end{equation}
so that the relevant Fourier modes are outside the Hubble radius and
photons and baryons are in the tight-coupling regime. We select the
regular growing adiabatic mode. Unlike models in which the scalar field
is initially frozen, the constant-$\epsilon$ background considered here
has
\begin{equation}
\bar{\phi}'
=
\frac{3\epsilon}{4\sigma}\mathcal{H}
\label{eq:initial-background-phi}
\end{equation}
for $\epsilon\neq0$ and $\sigma\neq0$. The scalar field perturbation must
therefore be included consistently in the adiabatic mode.

For any two components $i$ and $j$, adiabaticity requires
\begin{equation}
\frac{\delta\rho_i}{\bar{\rho}_i'}
=
\frac{\delta\rho_j}{\bar{\rho}_j'}.
\label{eq:general-adiabatic-condition}
\end{equation}
Since
\begin{equation}
\bar{\rho}_{\gamma}'
=
-(4-\epsilon)\mathcal{H}\bar{\rho}_{\gamma},
\end{equation}
whereas the uncoupled neutrino, baryon and CDM components
obey their standard background equations, the density perturbations
satisfy
\begin{equation}
\frac{\delta_\gamma}{4-\epsilon}
=
\frac{\delta_\nu}{4}
=
\frac{\delta_b}{3}
=
\frac{\delta_c}{3}.
\label{eq:adiabatic-density-relations}
\end{equation}

Adiabaticity between the scalar trajectory and the photon fluid requires
\begin{equation}
\frac{\delta\phi}{\bar{\phi}'}
=
\frac{\delta\rho_\gamma}{\bar{\rho}_\gamma'}
=
-\frac{\delta_\gamma}
{(4-\epsilon)\mathcal{H}}.
\label{eq:scalar-adiabatic-condition}
\end{equation}
Using equation~(\ref{eq:initial-background-phi}), this gives
\begin{equation}
\delta\phi_i
=
-\frac{3\epsilon}
{4\sigma(4-\epsilon)}
\delta_{\gamma i}.
\label{eq:delta-phi-initial-relation}
\end{equation}
Writing the regular synchronous gauge growing mode as
\begin{equation}
h=C(k\tau)^2,
\label{eq:initial-h}
\end{equation}
and neglecting velocity terms at leading super-Hubble order,
equations~\eqref{eq:photon-continuity-coupled} and
\eqref{eq:delta-phi-initial-relation} give
\begin{equation}
\delta_{\gamma i}
=
-\frac{4-\epsilon}{6}h_i.
\label{eq:initial-delta-gamma}
\end{equation}
The remaining density perturbations are therefore
\begin{equation}
\delta_{\nu i}
=
-\frac{2}{3}h_i,
\qquad
\delta_{b i}
=
\delta_{c i}
=
-\frac{1}{2}h_i,
\label{eq:initial-standard-species}
\end{equation}
while the scalar perturbation is
\begin{equation}
\delta\phi_i
=
\frac{\epsilon}{8\sigma}h_i.
\label{eq:initial-delta-phi}
\end{equation}
These expressions specify the leading regular adiabatic density and
scalar field initial conditions for the constant-$\epsilon$ branch.

The initial derivative of the scalar perturbation follows directly from
equations~(\ref{eq:initial-h}) and~(\ref{eq:initial-delta-phi}),
\begin{equation}
\delta\phi_i'
=
\frac{\epsilon}{8\sigma}h_i'
=
\frac{\epsilon}{4\sigma}Ck^2\tau_i.
\label{eq:initial-delta-phi-prime}
\end{equation}
The residual synchronous gauge freedom is fixed by CDM
rest frame,
\begin{equation}
\theta_{c i}=0,
\label{eq:initial-cdm-velocity}
\end{equation}
and tight coupling imposes
\begin{equation}
\theta_{\gamma i}=\theta_{b i}.
\label{eq:initial-tight-coupling}
\end{equation}
The remaining metric, velocity and collisionless radiation variables are
obtained from the regular adiabatic power series solution in $k\tau$
using the modified photon and scalar sources above. Their leading
regularity orders are
\begin{align}
\eta_i
&=
\eta_0+\mathcal{O}\!\left[(k\tau_i)^2\right],
\nonumber\\
\theta_{\gamma i}=\theta_{b i},
\ \theta_{\nu i}
&=
\mathcal{O}\!\left[k(k\tau_i)^3\right],
\nonumber\\
\Pi_{\nu i}
&=
\mathcal{O}\!\left[(k\tau_i)^2\right],
\qquad
F_{\nu 3,i}
=
\mathcal{O}\!\left[(k\tau_i)^3\right],
\label{eq:initial-regular-orders}
\end{align}
with $F_{\nu\ell,i}=\mathcal{O}[(k\tau_i)^\ell]$ for
$\ell\geq3$. During tight coupling the photon shear is not an
independent initial variable: $\Pi_{\gamma i}$ is supplied by the
standard tight-coupling relation while the higher photon temperature
and polarization multipoles are initialized with their standard
tight-coupling values. The coefficients of $\eta_i$, the common
photon--baryon velocity, the neutrino velocity, shear and higher
multipoles are calculated by the regular adiabatic initialization
already used in CLASS after replacing the photon density and
scalar field entries by equations~\eqref{eq:initial-delta-gamma},
\eqref{eq:initial-delta-phi} and
\eqref{eq:initial-delta-phi-prime}
\cite{ma_bert,class}. 
For $\epsilon=0$, equations~\eqref{eq:initial-delta-gamma},
\eqref{eq:initial-delta-phi} and
\eqref{eq:initial-delta-phi-prime} reduce respectively to the standard
photon density condition and vanishing scalar perturbation and
derivative. The complete initial vector therefore reduces continuously
to the standard CLASS regular adiabatic mode.

\paragraph{Data and code availability.} The observational data used in this work are publicly available through the corresponding survey and likelihood releases. The modified CLASS and MontePython configuration files, plotting scripts and numerical chains can be made available by the author upon reasonable request.\\\\
\textbf{Acknowledgement:} This work is supported by Shahi Rajaee Teacher Training Univesity under the grant no. 23905/380001.\\\\\\

\end{document}